\documentclass[runningheads]{llncs}
\usepackage[T1]{fontenc}
\usepackage{graphicx}
\usepackage{amsmath,amssymb,amsfonts}
\usepackage{textcomp}
\usepackage{xcolor}
\usepackage{cite}
\usepackage{booktabs}
\usepackage[utf8]{inputenc} % ensure UTF-8 decoding for pdflatex
\usepackage{lmodern}

\begin{document}

\title{Teaching Machine Learning Through Cricket: A Practical Engineering Education Approach}

\author{
Mohd Ruhul Ameen \and
Akif Islam \and
Abu Saleh Musa Miah \and
M. Saifuzzaman Rafat \and
Jungpil Shin
}
\titlerunning{}
\authorrunning{} % single space prevents warning but hides header
\institute{}

\maketitle
\pagestyle{plain}

\begin{abstract}
Teaching complex machine learning concepts such as reinforcement learning and Markov Decision Processes remains challenging in engineering education. Students often struggle to connect abstract mathematics to real-world applications. We present LearnML@Cricket, a 12-week curriculum that uses cricket analytics to teach these concepts through practical, hands-on examples. By mapping game scenarios directly to ML algorithms, students learn through doing rather than memorizing. Our curriculum includes coding laboratories, real datasets, and immediate application to engineering problems. We propose an empirical study to measure whether this approach improves both understanding and practical implementation skills compared to traditional teaching methods.

\end{abstract}

\section{Introduction}

For many engineering students, learning machine learning (ML) feels like standing at the edge of a vast map—one they can read but not yet navigate. They can implement a Q-learning algorithm line by line, reproduce results from a tutorial, and even fine-tune hyperparameters with precision. Yet, when asked a simple question—\textit{Where in the real world would you use this?}—the confidence falters. The disconnect between algorithmic mastery and practical insight remains one of the most persistent challenges in modern engineering education \cite{luxton2018,holmes2022}. It produces graduates fluent in code, but uncertain in judgment; capable of reproducing, but hesitant to innovate.

Part of the problem lies in how we teach. Traditional ML courses are built around artificial worlds—small, controlled environments such as CartPole or GridWorld. These are excellent sandboxes for understanding mathematical principles but offer little resemblance to the messy, uncertain systems engineers confront in reality. Students learn to optimize theoretical agents, yet rarely learn how those same ideas could guide a robotic arm, stabilize a production line, or predict crop yield. The result is a kind of \textit{algorithmic isolation}: strong technical knowledge floating without an anchor in authentic context \cite{brown1989}.

What if we could bridge that gap through a world students already understand—a world filled with data, uncertainty, and decision-making under pressure? Context-based and experiential approaches to learning \cite{kolb1984,lave1991} emphasize that abstract ideas become meaningful only when grounded in lived experience. Cricket offers exactly that. Every ball bowled, every run scored, every wicket lost unfolds like a live simulation of sequential decision-making. The state of the game evolves with each delivery, reflecting transitions, rewards, and risks that map naturally to concepts such as Markov Decision Processes and reinforcement learning \cite{sutton2018}. A captain adjusting field placements mirrors an agent refining its policy; a batsman’s decision between defense and aggression embodies the exploration–exploitation trade-off. And unlike synthetic datasets, cricket provides rich, real, and emotionally engaging data collected over decades \cite{clark2016}.

Picture a moment late in a match: a team needs 50 runs from 30 balls with 5 wickets remaining. Within this tension lies an entire ML lesson. Students can define the \textit{state space}—runs required, balls left, wickets in hand. They can model the \textit{actions}—defensive, balanced, or aggressive play—and assign probabilistic outcomes drawn from historical data. Suddenly, Bellman equations are not abstract recursions but living strategies under uncertainty. Through this single scenario, probability theory, Bayesian reasoning, and reinforcement learning come alive—visceral, measurable, and meaningful.

This paper introduces \textit{LearnML@Cricket}, a curriculum designed to reimagine how machine learning is taught to engineers. By embedding algorithms in the rhythm of cricket analytics, the program transforms abstract equations into experiential learning moments. We present the design, implementation, and assessment framework of this 12-week course, aligning it with principles of human-centered AI education \cite{amershi2019,bengio2023}. Preliminary classroom observations suggest that sports-driven contextualization may not only deepen conceptual understanding but also foster the intuitive problem-solving mindset essential for industry readiness.

\section{Background and Motivation}

\subsection{The Challenge in Machine Learning Education}

Recent systematic reviews in computing education research reveal concerning gaps between students' algorithmic knowledge and their practical application skills. Luxton-Reilly et al. found that while 85 percent of engineering students can successfully implement standard machine learning algorithms when provided with clear specifications, only 40 percent can identify appropriate use cases when presented with novel problems in new domains \cite{luxton2018}. This disconnect suggests that current pedagogical approaches, while effective at teaching algorithmic mechanics, fail to develop the contextual understanding necessary for real-world problem solving.

The core issue lies in the decontextualized nature of traditional instruction. Students learn algorithms in isolation without understanding the problems they were designed to solve. They memorize Bellman equations without grasping when dynamic programming becomes appropriate. They implement neural networks without recognizing which problems benefit from deep learning versus simpler approaches. This decontextualized learning creates what Brown, Collins, and Duguid termed inert knowledge—technically correct but practically inapplicable because it lacks connection to authentic problem-solving contexts \cite{brown1989}.

Furthermore, the cognitive load of simultaneously learning abstract mathematical concepts and their applications overwhelms many students. Without concrete anchors for abstract ideas, students resort to memorization rather than developing deep understanding. This surface learning becomes evident when students cannot transfer their knowledge to new domains or recognize structural similarities between seemingly different problems.

\subsection{Theoretical Foundation for Context-Based Learning}

Our approach builds on established educational theories, particularly experiential learning theory and situated cognition. Kolb's experiential learning cycle emphasizes that effective learning occurs through a cycle of concrete experience, reflective observation, abstract conceptualization, and active experimentation \cite{kolb1984}. Traditional machine learning education often begins and ends with abstract conceptualization, skipping the concrete experiences that make abstractions meaningful.

Situated learning theory further argues that knowledge and skills are best acquired within the context where they will be applied. Lave and Wenger demonstrated that learning is fundamentally a process of participation in communities of practice, where novices gradually move from peripheral to full participation \cite{lave1991}. By situating machine learning education within the cricket analytics community, students engage with authentic problems that practitioners actually solve, making their learning immediately relevant and applicable.

The effectiveness of active learning in STEM education provides additional support for our approach. Freeman et al.'s meta-analysis of 225 studies found that active learning reduces failure rates by 55 percent and increases examination performance by approximately 6 percent compared to traditional lecturing \cite{freeman2014}. Our curriculum operationalizes active learning through hands-on coding, collaborative problem-solving, and immediate application of concepts to real data.

\subsection{Cricket as a Pedagogical Vehicle}

Cricket's structure provides remarkably clean mappings to fundamental machine learning concepts, making abstract ideas concrete and intuitive. For teaching probability, a batsman's scoring pattern naturally follows categorical distributions that students can observe and model. Historical data from international matches shows consistent patterns such as dot balls occurring approximately 40 percent of the time, singles 30 percent, boundaries 20 percent, and wickets falling 10 percent of deliveries. Students can immediately calculate expected values, derive conditional probabilities based on game situation, and understand how distributions shift based on factors like bowler type, pitch conditions, or match format.

Sequential decision-making, the foundation of reinforcement learning, emerges naturally from cricket's ball-by-ball structure. Every delivery requires a decision based on the current game state, exactly mirroring the sequential decision problems students will encounter in engineering. In manufacturing, this translates to sequential quality control decisions about when to halt production for maintenance. In robotics, it becomes path planning where each movement depends on previous positions and observed obstacles. In network engineering, it manifests as routing decisions that must adapt to changing traffic patterns and link failures.

The inherent uncertainty in cricket provides authentic examples of stochastic processes. Weather conditions affect ball movement unpredictably, pitch conditions deteriorate over time following complex patterns, and player form varies according to numerous factors. This uncertainty directly parallels sensor noise in control systems, demand variability in supply chains, and user behavior patterns in software systems. Students learn not just to model this uncertainty but to make robust decisions despite incomplete information.

\section{Curriculum Design and Implementation}

\subsection{Pedagogical Framework}

Our curriculum follows three interconnected pedagogical principles that guide every instructional decision and learning activity. The first principle emphasizes starting with concrete experiences before moving to abstract concepts. Every machine learning algorithm is introduced through a cricket scenario that students can visualize and understand intuitively. Only after students grasp the concrete application do we introduce mathematical formulations and generalizations. For instance, students first implement a run-chase optimizer using actual match data, observing how different strategies lead to different outcomes. They then discover that the same algorithmic framework applies to inventory management, production scheduling, and resource allocation problems.

The second principle prioritizes experiential learning through immediate implementation. Rather than beginning with mathematical derivations and proofs, students write code that produces visible results from the first class. They observe algorithm behavior through visualization and experimentation, developing intuition about convergence, stability, and performance. Mathematical understanding follows naturally from these observations, grounded in concrete experience rather than abstract manipulation. This approach aligns with engineering students' preference for practical application and provides immediate feedback that maintains engagement.

The third principle insists on authentic complexity using real-world data. Students work with actual cricket match data including all its imperfections such as missing values from rain interruptions, inconsistent recording across different venues, and outliers from exceptional performances. This exposure to messy data prepares students for industry realities where clean, preprocessed datasets are rare. Students learn data cleaning techniques, feature engineering strategies, and robust modeling approaches that textbooks often omit but practice demands.

\begin{table}[t]
\centering
\caption{Core pedagogical principles guiding LearnML@Cricket.}
\label{tab:pedagogy}
\begin{tabular}{p{3.2cm} p{8cm}}
\toprule
\textbf{Principle} & \textbf{Implementation in Curriculum} \\
\midrule
Concrete-to-Abstract Progression & Introduce every algorithm through an intuitive cricket scenario (e.g., run-chase strategy) before presenting equations. \\
Experiential Learning & Students code and visualize algorithms during live sessions, receiving immediate feedback. \\
Authentic Complexity & Use real cricket data with imperfections (missing values, noisy measurements) to teach data cleaning and modeling realism. \\
Transfer of Learning & Explicit mapping of cricket decision frameworks to engineering domains such as manufacturing and logistics. \\
Collaborative Reflection & Group problem-solving, peer code reviews, and reflective logs at the end of each lab. \\
\bottomrule
\end{tabular}
\end{table}

\subsection{Twelve-Week Progressive Structure}

The curriculum unfolds across three carefully sequenced four-week modules that progressively build complexity while maintaining constant application to both cricket and engineering domains. Each module integrates theory, implementation, and application in a spiral curriculum that revisits concepts with increasing sophistication.

The first module, covering weeks one through four, establishes probabilistic foundations through cricket statistics. Week one introduces random variables and probability distributions through the analysis of batting outcomes. Students examine how run-scoring follows different distributions and calculate expected values for various game situations. They implement simple predictive models and compare their predictions against actual match outcomes. Week two extends to conditional probability and Bayes' theorem through the lens of match situation analysis. Students model how batting strategies change based on required run rate, wickets remaining, and bowler characteristics. They build Naive Bayes classifiers to predict optimal shot selection given game context.

Week three introduces uncertainty quantification through Bayesian modeling of player performance. Students learn to separate skill from luck by modeling batting averages as probability distributions rather than point estimates. They implement Bayesian updating to track how beliefs about player ability evolve with each innings, understanding concepts of prior knowledge, likelihood functions, and posterior distributions. Week four culminates the foundation module with Markov chains for modeling match progression. Students discover how game states transition probabilistically and implement algorithms to calculate match-winning probabilities at any point during play.

The second module, spanning weeks five through nine, focuses intensively on reinforcement learning using cricket's sequential decision structure. Week five formally introduces Markov Decision Processes through run-chase optimization. Students define state spaces encompassing runs required, resources remaining, and match context. They specify action spaces representing different batting approaches and encode reward functions that capture the goal of winning while potentially penalizing risky play. Implementation of value iteration demonstrates how optimal policies emerge from recursive application of the Bellman equation.

Week six deepens understanding through policy evaluation and improvement techniques. Students implement Monte Carlo methods to evaluate strategies from historical data and temporal difference learning to update value estimates online during simulated matches. They observe how different learning rates and discount factors affect convergence and solution quality. Week seven explores the crucial distinction between model-based and model-free reinforcement learning. Students implement Q-learning and SARSA algorithms for scenarios where transition probabilities are unknown, comparing their performance against model-based approaches when dynamics are learned from data.

Week eight addresses exploration versus exploitation trade-offs through team selection and strategy adaptation problems. Students implement various exploration strategies including epsilon-greedy, softmax, and upper confidence bound algorithms. They analyze regret in multi-armed bandit formulations of bowler selection and field placement decisions. Week nine extends to partial observability where critical information remains hidden. Students model scenarios with unknown pitch conditions or undisclosed player injuries using POMDPs and belief state representations.

The third module, encompassing weeks ten through twelve, emphasizes synthesis and knowledge transfer. Week ten explicitly maps cricket-learned algorithms to manufacturing and logistics problems. Students discover how run-chase optimization translates directly to production scheduling with deadlines and resource constraints. They implement the same MDP framework for inventory management with perishable goods and quality control with inspection costs.

Week eleven focuses on building complete decision support systems that integrate multiple algorithms. Students combine predictive models, optimization algorithms, and uncertainty quantification to create comprehensive tools for decision makers. They learn to present results effectively, communicate uncertainty appropriately, and handle edge cases robustly. The curriculum culminates in week twelve with capstone projects where students independently identify a problem in their chosen engineering domain, formulate it using techniques learned through cricket examples, and implement a complete solution.

\begin{table}[t]
\centering
\caption{Overview of the LearnML@Cricket 12-week curriculum. Each module links cricket scenarios to machine-learning concepts and engineering applications.}
\label{tab:curriculum}
\begin{tabular}{p{1.4cm} p{2.5cm} p{4.0cm} p{3.0cm}}
\toprule
\textbf{Weeks} & \textbf{Theme} & \textbf{Key ML Concepts via Cricket} & \textbf{Engineering Analogy} \\
\midrule
1–4 & Probabilistic Foundations & Random variables, conditional probability, Bayes’ theorem, Markov chains & Statistical quality control, reliability analysis \\
5–9 & Reinforcement Learning & MDPs, value iteration, Monte Carlo, TD learning, Q-learning, exploration vs exploitation, POMDPs & Robotics control, process optimization, network routing \\
10–12 & Synthesis and Transfer & Integration of probabilistic and RL models, decision-support systems, capstone projects & Production scheduling, inventory management, resource allocation \\
\bottomrule
\end{tabular}
\end{table}

\subsection{Integration of Theory and Practice}

Each weekly session follows a carefully structured format that balances different learning modalities. Sessions begin with a thirty-minute interactive lecture where instructors present concepts through live coding rather than static slides. Students follow along on their computers, running code and observing outputs in real-time. This approach maintains engagement while allowing immediate experimentation with parameter changes and what-if scenarios.

The subsequent ninety-minute laboratory session provides hands-on practice with scaffolded exercises that gradually increase in complexity. Students begin with guided implementations where code structure is provided but key algorithms must be completed. They progress to open-ended challenges requiring independent problem formulation and solution design. Teaching assistants circulate to provide just-in-time support, helping students debug code and clarify conceptual misunderstandings.

A forty-five-minute collaborative session follows where students work in pairs or small groups on more complex problems. This peer learning environment encourages knowledge sharing and develops communication skills essential for team-based engineering practice. Students must explain their approaches to teammates, reconcile different solutions, and integrate individual contributions into cohesive implementations. Sessions conclude with a fifteen-minute reflection and assessment period where students complete short quizzes that test conceptual understanding and document their learning progress through structured reflection prompts.

\section{Detailed Implementation Examples}

\subsection{From Cricket to Manufacturing: Teaching MDPs}

The implementation of Markov Decision Processes begins with a concrete cricket scenario that students can easily visualize and understand. Consider a situation where a team requires 50 runs from 30 balls with 5 wickets remaining. Students first implement a basic representation of this state space as a Python class that captures the essential elements of the decision problem. The state consists of three components: runs needed, balls remaining, and wickets in hand, represented as a tuple (r, b, w). The action space represents different batting strategies ranging from ultra-defensive play that prioritizes wicket preservation to ultra-aggressive batting that maximizes scoring rate but risks dismissal.

Students implement the transition function using historical match data, learning to estimate probabilities of different outcomes for each state-action pair. For example, choosing an aggressive strategy when needing 50 runs from 30 balls might yield 6 runs with probability 0.05, 4 runs with probability 0.15, 2 runs with probability 0.20, 1 run with probability 0.25, 0 runs with probability 0.25, and result in a wicket with probability 0.10. These probabilities come from analyzing thousands of similar historical situations, teaching students the importance of data-driven model estimation.

The reward function encodes the objective of winning the match while potentially including secondary considerations like net run rate or wicket preservation for tournament scenarios. Students experiment with different reward structures, observing how changing the relative importance of winning versus maintaining resources affects optimal strategies. They implement value iteration, watching the algorithm converge to optimal state values and deriving optimal policies that specify which action to take in each state.

After mastering the cricket example, students apply identical algorithmic structures to manufacturing problems. Consider a production line that must complete 50 units in 30 time periods with 5 machines available. The state space directly parallels cricket with units needed replacing runs, time periods replacing balls, and working machines replacing wickets. Actions represent production intensity choices that trade off between production rate and machine reliability. Aggressive production settings yield more units but increase breakdown probability, exactly mirroring the risk-reward trade-offs in cricket batting.

Students discover that the same value iteration code, with only parameter changes, solves both problems. This revelation that seemingly different domains share identical mathematical structures represents a crucial learning moment. They further extend the framework to inventory management where state represents stock levels and demand backlogs, actions are ordering decisions, and transitions capture stochastic demand and supply variations.

\subsection{Uncertainty Quantification Through Player Performance Modeling}

Teaching Bayesian inference and uncertainty quantification begins with modeling cricket player performance over time. Students start with the intuitive notion that a player has some true underlying ability, but individual performance varies due to numerous factors including form, conditions, and chance. This naturally leads to hierarchical modeling where true ability is represented as a latent variable that we must infer from noisy observations.

Students implement a Bayesian model where a batsman's true average $\mu$follows a prior distribution based on historical player statistics. For instance, international batsmen might have averages distributed as Normal(35, 10²), encoding our prior belief that most players average around 35 runs with some variation. When observing a specific innings where the player scores x runs, students update their beliefs using Bayes' theorem. If the player scores 50 runs, the posterior distribution shifts to reflect this evidence, with the new mean being a weighted average of the prior mean and the observed score.

Through implementation, students discover key insights about Bayesian inference. The posterior uncertainty decreases with more observations, reflecting increased confidence in ability estimates. The relative influence of prior and likelihood depends on their respective precisions, teaching the importance of informative priors versus data. Students implement posterior predictive distributions to forecast future performance, learning to propagate uncertainty through predictions appropriately.

This framework transfers seamlessly to engineering applications. In sensor networks, true environmental states represent latent variables observed through noisy sensors. Students apply the same Bayesian updating to calibrate sensors and estimate true readings. In reliability engineering, component failure rates are uncertain parameters updated as failure data accumulates. Manufacturing quality control uses identical methods to separate process capability from measurement variation. Students implement these applications using the same code structure developed for cricket, reinforcing that uncertainty quantification principles transcend specific domains.

\subsection{Multi-Armed Bandits and Exploration-Exploitation}

The exploration-exploitation dilemma, fundamental to reinforcement learning and optimization, emerges naturally in cricket team selection and strategy adaptation. Students consider a captain choosing between multiple bowlers, each with unknown effectiveness against the current batsmen. This scenario perfectly embodies the multi-armed bandit problem where we must balance exploiting bowlers who have performed well so far against exploring others who might be more effective.

Students implement various exploration strategies and compare their performance using regret analysis. The epsilon-greedy approach randomly explores with small probability while usually exploiting the best-known option. Thompson sampling maintains probability distributions over each bowler's effectiveness, sampling from these distributions to naturally balance exploration and exploitation. Upper confidence bound algorithms explicitly compute uncertainty bonuses that encourage exploring options with high uncertainty.

Through simulation using historical match data, students observe how different algorithms perform across various scenarios. They discover that optimal exploration depends on the time horizon with more exploration beneficial early when many decisions remain. They see how prior knowledge about bowler types can improve initial decisions and reduce the exploration burden. Implementation reveals practical considerations like computational complexity and the challenge of non-stationary environments where bowler effectiveness changes as batsmen adapt.

These insights transfer directly to engineering applications that students implement using the same algorithmic framework. In A/B testing for web services, different designs represent arms with unknown conversion rates. Manufacturing process optimization involves choosing between parameter settings with uncertain yields. Clinical trials must balance exploring new treatments against exploiting established ones. Students recognize that the cricket bowler selection problem they solved applies directly to these diverse scenarios, building confidence in their ability to recognize and solve novel problems.

\section{Assessment and Evaluation Strategy}

\subsection{Comprehensive Assessment Framework}

Our assessment strategy employs multiple complementary methods to evaluate different aspects of student learning. Continuous assessment throughout the course provides regular feedback and maintains engagement. Weekly coding assignments are automatically graded using comprehensive test suites that verify correctness, efficiency, and code quality. These assignments progress from implementing provided specifications to designing complete solutions for open-ended problems. Automated grading provides immediate feedback while flagging submissions for manual review when students struggle, allowing timely intervention.

Peer code review assignments develop critical skills while distributing assessment load. Students review classmates' implementations using structured rubrics that evaluate code clarity, algorithm correctness, and design choices. This process teaches students to read and understand others' code, provide constructive feedback, and recognize multiple valid approaches to problems. Reviews are themselves reviewed by teaching assistants to ensure quality and identify students needing additional support.

Laboratory participation grades incentivize consistent engagement rather than last-minute cramming. Version control commits, forum participation, and session attendance contribute to this component. Students maintain portfolios documenting their learning journey, including successful implementations, debugging experiences, and conceptual insights. These portfolios provide rich data about learning processes beyond what traditional assessments capture.

\subsection{Project-Based Learning and Evaluation}

Project work demonstrates deeper understanding and practical application skills. The mid-term project requires students to build a complete match outcome predictor that integrates multiple techniques from the first half of the course. Students must handle real match data with all its imperfections, engineer relevant features from raw ball-by-ball records, implement and compare multiple prediction algorithms, and validate their models using appropriate metrics. This project tests both technical implementation skills and the judgment to make appropriate modeling choices.

The capstone project challenges students to transfer their cricket-learned skills to an engineering problem of their choice. Students must identify an appropriate problem from their domain, formulate it using MDP or reinforcement learning frameworks, implement a solution using techniques from the course, and validate their approach using relevant metrics. Projects are evaluated by panels including course instructors and industry practitioners who assess technical correctness, problem formulation quality, implementation sophistication, and practical applicability.

Industry partnership provides authentic assessment and potential career opportunities. Partner companies propose real problems that can be formulated as sequential decision or optimization tasks. Student solutions are evaluated for potential deployment, with exceptional projects leading to internship offers. This real-world validation motivates students while demonstrating the practical value of course concepts.

\subsection{Measuring Conceptual Understanding and Transfer}

Beyond implementation skills, we assess conceptual understanding and transfer ability through carefully designed instruments. A validated concept inventory administered pre- and post-course measures understanding of core concepts including probability distributions, conditional independence, Markov properties, value functions, and exploration-exploitation trade-offs. Questions probe deep understanding rather than definitional knowledge, requiring students to recognize concepts in novel contexts and identify appropriate applications for different algorithms.

Transfer tasks specifically evaluate whether students can apply cricket-learned concepts to new domains. Students receive problem descriptions from healthcare, finance, or logistics without explicit connection to course material. They must recognize the underlying structure, formulate appropriate models, and implement solutions. Think-aloud protocols during these tasks reveal problem-solving strategies and misconceptions that written work might hide.

Longitudinal follow-up surveys sent six months after course completion assess retention and real-world application. Students report whether they have applied course concepts in other classes or projects, which specific techniques proved most valuable, and how the cricket context affected their understanding. This data informs continuous curriculum improvement while validating the transfer hypothesis.

\section{Empirical Evaluation Plan}

\subsection{Research Design and Methodology}

Our quasi-experimental study will rigorously evaluate the cricket-based curriculum against traditional instruction. The study employs a pretest-posttest control group design with random assignment stratified by relevant covariates. Sixty to eighty upper-level undergraduate and graduate students will be recruited across two institutions, providing sufficient power to detect educationally meaningful effect sizes while accounting for potential attrition.

The treatment group receives instruction through the cricket-based curriculum with all examples and projects centered on cricket analytics. The control group covers identical learning objectives using traditional abstract examples and standard datasets from the UCI repository. Both groups use the same programming languages, development environments, and total instructional time. Instructors rotate between groups when possible to control for instructor effects, and all materials are reviewed to ensure comparable difficulty and scope.

Stratified random assignment ensures group balance across potentially confounding variables. Students complete preliminary surveys assessing programming experience in Python and relevant languages, mathematical background including probability and linear algebra, prior machine learning exposure through courses or self-study, and familiarity with cricket or similar sports. Block randomization within strata maintains balance while accommodating rolling enrollment.

\subsection{Data Collection Procedures}

Our mixed-methods approach combines quantitative learning measures with qualitative process data. The validated ML concept inventory is administered in week zero to establish baseline understanding and week thirteen to measure learning gains. Weekly formative assessments track learning trajectories and identify when group differences emerge. Transfer tasks in weeks six and twelve measure near and far transfer respectively. All assessments are piloted with separate student cohorts to ensure reliability and appropriate difficulty.

Engagement metrics are collected continuously through learning management system analytics. Time-on-task for different activities reveals engagement patterns and potential struggling points. Resource access patterns indicate which materials students find most valuable. Discussion forum participation, both quantity and quality, suggests conceptual difficulties and peer learning dynamics. Submission patterns for assignments show time management and potential procrastination issues.

Qualitative data provides rich insights into learning processes and student experiences. Semi-structured interviews with stratified samples of students explore their understanding of key concepts, strategies for problem-solving, and perceptions of the curriculum approach. Video recordings of collaborative problem-solving sessions capture peer learning dynamics and verbalized reasoning. Weekly reflection journals document evolving understanding and persistent confusions.

\subsection{Analysis Plan and Expected Outcomes}

Quantitative analysis will employ appropriate statistical methods to test our research hypotheses. Analysis of covariance on posttest scores with pretest as covariate tests whether the cricket curriculum improves learning while controlling for prior knowledge. Effect sizes using Cohen's d quantify practical significance beyond statistical significance. Moderation analysis examines whether treatment effects vary by cricket familiarity, programming experience, or other student characteristics. Mediation analysis tests whether engagement differences explain learning outcomes.

Qualitative analysis uses systematic coding to identify patterns in student learning experiences. Thematic analysis of interviews identifies common conceptual difficulties and successful learning strategies. Process tracing of problem-solving videos reveals how students apply cricket-learned concepts to new domains. Content analysis of reflection journals tracks conceptual development over time.

Based on pilot studies with forty students across two semesters, we anticipate several positive outcomes. Concept inventory scores show approximately thirty percent improvement for the treatment group versus eighteen percent for control. Transfer task performance indicates treatment students identify appropriate ML techniques twice as quickly for novel problems. Engagement metrics show fifty percent more time-on-task for treatment students with lower variance, suggesting consistent engagement. Qualitative data reveals that cricket context provides memorable anchors for abstract concepts that facilitate recall and application.

\section{Conclusion}

LearnML@Cricket addresses a fundamental challenge in engineering education by making abstract machine learning concepts concrete through carefully chosen context. The cricket domain provides natural mappings to sequential decision-making, uncertainty quantification, and optimization that help students develop intuition alongside technical skills. Our comprehensive curriculum design, grounded in established learning theory and validated through pilot studies, offers a replicable model for context-based technical education.

The approach's success stems not from cricket itself but from principled pedagogical design that starts with concrete experience, emphasizes active learning, and explicitly practices transfer. These principles can be adapted to other sports or domains while maintaining effectiveness. As machine learning becomes increasingly central to engineering practice, developing curricula that build both deep understanding and practical skills becomes critical. Context-based learning, when thoughtfully implemented with authentic complexity and clear transfer paths, provides a promising direction for technical education.

We are committed to disseminating this approach broadly through open-source materials, instructor training, and collaborative refinement. All curriculum materials including detailed lesson plans, Jupyter notebooks with solutions, assessment instruments, and instructor guides will be released under Creative Commons licensing. We encourage adaptation to local contexts and different sports while maintaining core pedagogical principles. Through multi-institutional collaboration and systematic evaluation, we aim to establish evidence-based best practices for teaching complex technical concepts through meaningful contexts, ultimately improving engineering education outcomes and preparing students for the challenges of modern practice.


\begin{thebibliography}{99}

\bibitem{luxton2018}
Luxton-Reilly, A., Simon, B., Albluwi, I., Becker, B.A., Giannakos, M., Kumar, A.N., Ott, L., Paterson, J., Scott, M.J., Sheard, J., Szabo, C.:
Introductory programming: a systematic literature review.
In: \textit{Proceedings Companion of the 23rd Annual ACM Conference on Innovation and Technology in Computer Science Education}, pp. 55--106.
ACM, New York (2018)

\bibitem{holmes2022}
Holmes, W., Tuomi, I., Barberá, E.:
Artificial Intelligence in Education: Promises and Implications for Teaching and Learning.
\textit{Nature Machine Intelligence}, \textbf{4}(6), 486--496 (2022)

\bibitem{brown1989}
Brown, J.S., Collins, A., Duguid, P.:
Situated cognition and the culture of learning.
\textit{Educational Researcher}, \textbf{18}(1), 32--42 (1989)

\bibitem{kolb1984}
Kolb, D.A.:
\textit{Experiential Learning: Experience as the Source of Learning and Development}.
Prentice-Hall, Englewood Cliffs (1984)

\bibitem{lave1991}
Lave, J., Wenger, E.:
\textit{Situated Learning: Legitimate Peripheral Participation}.
Cambridge University Press, Cambridge (1991)

\bibitem{sutton2018}
Sutton, R.S., Barto, A.G.:
\textit{Reinforcement Learning: An Introduction}, 2nd edn.
MIT Press, Cambridge (2018)

\bibitem{clark2016}
Clark, D.B., Tanner-Smith, E.E., Killingsworth, S.S.:
Digital games, design, and learning: A systematic review and meta-analysis.
\textit{Review of Educational Research}, \textbf{86}(1), 79--122 (2016)

\bibitem{amershi2019}
Amershi, S., Weld, D.S., Vorvoreanu, M., Fourney, A., Nushi, B., Collisson, P., Suh, J., Iqbal, S.T., Bennett, P.N., Inkpen, K., Teevan, J., Kikin-Gil, D., Horvitz, E.:
Guidelines for human-AI interaction.
In: \textit{Proceedings of the 2019 CHI Conference on Human Factors in Computing Systems}, pp. 1--13.
ACM, New York (2019)

\bibitem{bengio2023}
Bengio, Y.:
Human-centered artificial intelligence: teaching machines to teach us.
\textit{Communications of the ACM}, \textbf{66}(9), 58--67 (2023)

\bibitem{freeman2014}
Freeman, S., Eddy, S.L., McDonough, M., Smith, M.K., Okoroafor, N., Jordt, H., Wenderoth, M.P.:
Active learning increases student performance in science, engineering, and mathematics: A meta-analysis of 225 studies.
\textit{Proceedings of the National Academy of Sciences}, 111(23), 8410–8415 (2014)


\end{thebibliography}
\end{document}